\documentclass[12pt]{article}

\usepackage{graphicx}

\usepackage{amsmath,amssymb}

\title{Optomechanical self-structuring in cold atomic gases}

\author
{Guillaume Labeyrie$^{1\ast}$, Enrico Tesio$^{2}$, Pedro M. Gomes$^{2}$, \\
Gian-Luca Oppo$^{2}$, William J. Firth$^{2}$, Gordon R. M. Robb$^{2}$, \\
Aidan S. Arnold$^{2}$, Robin Kaiser$^{1,2\ast}$, and Thorsten Ackemann$^{2\ast}$
\\
\normalsize{$^{1}$Institut Non Lin\'{e}aire de Nice, UMR 6618 CNRS, 1361 route des
Lucioles,} \\
\normalsize{06560 Valbonne, France}\\
\normalsize{$^{2}$SUPA and Department of Physics, University of Strathclyde,}\\
\normalsize{Glasgow G4 0NG, Scotland, UK}
}

\begin{document}

\maketitle

\textbf{The rapidly developing field of optomechanics aims at the combined control of optical and mechanical (solid-state or atomic) 
modes \cite{brennecke08,favero09,ritsch13}. In particular, laser cooled atoms have been used to exploit optomechanical coupling for self-organization in a variety of schemes where the accessible length scales are constrained by a combination of pump modes and those associated to a second imposed axis, typically a cavity axis
\cite{ritsch13,inouye99, domokos02, black03, voncube04, baumann10, greenberg11}.
Here, we consider a system with many spatial degrees of freedom around a single distinguished axis, in which two symmetries - rotations and translations in the plane orthogonal to the pump axis -  are spontaneously broken. We observe the simultaneous spatial structuring of the density of a cold atomic cloud and an optical pump beam. The resulting patterns have hexagonal symmetry. The experiment demonstrates the manipulation of matter by opto-mechanical self-assembly with adjustable length scales and can be potentially extended to quantum degenerate gases.}

In coupled light-atom systems, a pump field interacts with a cloud of cold atoms and a probe field (possibly a noise field) propagating along the same or a separate axis \cite{ritsch13, domokos02, black03, voncube04, greenberg11}. The interference pattern created by these two fields causes a bunching of the cold atoms due to dipole forces. The resulting density grating then scatters pump light into the probe mode, which amplifies it  and enhances the density grating, leading to instability and eventually sustained light output along the probe axis. If the sample does not consists of cold atoms but of macroscopic quantum matter like a Bose-Einstein condensate, the phenomenology is quite similar but the resulting transition has been interpreted as a quantum phase transition \cite{baumann10}. `Hybrid' systems, in which a density grating providing feedback is created via opticalmechanical forces imposed externally but the gain mechanism for the instability is due to four-wave mixing via internal states, found also considerable interest \cite{schilke12}.

Recent theoretical work predicted the spontaneous emergence of two-dimensional purely optomechanical structures in the plane transverse to a single pump axis \cite{greenberg11,saffman98,muradyan05,notegreenberg11,tesio12}, with optical diffraction providing a crucial additional mechanism for spatial coupling. Self-organized diffractive structures have been observed in hot atomic vapours for a number of years \cite{grynberg88,ackemann94,dawes05}, but these rely solely on variations of the internal states of the atoms (electronic or Zeeman states) without affecting the centre-of-mass motion of the atoms. The crucial new ingredient in cold atoms is the existence of macroscopic transport processes of matter due to dipole forces, leading to density structures of linear scatterers without the need for an intrinsic optical nonlinearity. The grating structure formed in the atomic density modifies linear light propagation so as to enhance the dipole forces. This feedback loop results in a beautiful example of spontaneous self-structuring in systems driven out of thermal equilibrium with broad relevance to natural sciences and technology also outside physics. Some form of material transport via convection (hydrodynamics), diffusion (chemistry and biology), or charge drift (gas discharges) is typically present in pattern-forming systems, but modulation of the overall matter density is neither the decisive driver nor the manifestation of self-organization. Hence, the self-structuring and manipulation of the density of matter demonstrated in this communication augments the variety of pattern forming systems. Interesting parallels may exist between optomechanical density patterns and self-structuring phenomena in a diverse range of systems including, e.g., plasmas with ponderomotive forces \cite{tesio13} and  colonies of bacteria displaying chemotaxis \cite{budrene95}.

\begin{figure}
\begin{center}
\includegraphics[width=160mm]{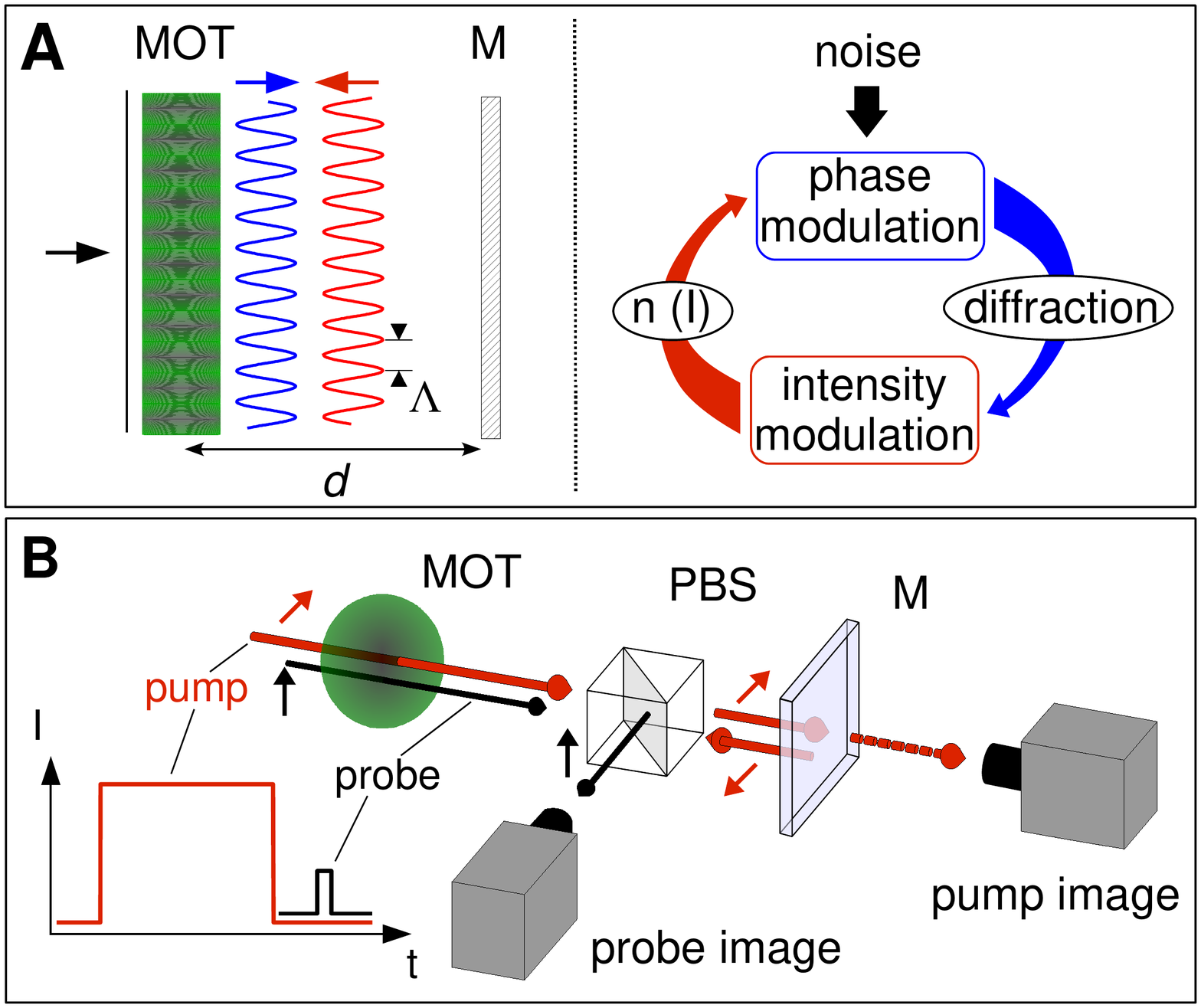}
\caption{(\textbf{A}) Mechanism for self-organization. A plane wave is incident on a cloud of cold atoms (MOT), and the transmitted wave is retro-reflected by a mirror (M) located at a distance $d$. In the presence of fluctuations in the cloud refractive index, the transmitted wavefront presents a transverse spatial modulation (blue curve). Diffraction during free-space propagation to the mirror and back to the sample converts the phase modulation to an intensity modulation (red curve). This intensity modulation induces a modulation of the cloud index, which can enhance the initial fluctuation. This positive feedback loop, sketched in the right panel, leads to a transverse instability and spontaneous appearance of patterns. (\textbf{B}) Experimental setup. A linearly-polarized pump beam passes through the cold atomic cloud, and is retro-reflected by the feedback mirror. The light transmitted by the mirror is sent to a CCD (pump image). A weak probe pulse of orthogonal polarization can also be sent through the cloud along the same path, after the pump pulse has been switched off, to probe transverse spatial ordering. The transmitted probe light is separated from the pump by a polarizing beam-splitter (PBS) and sent to another CCD (probe image).}
\label{fig1}
\end{center}
\end{figure}

An effective and conceptually simple scheme for optical pattern formation is the single mirror feedback arrangement illustrated in Fig.~\ref{fig1}\textbf{A}~\cite{firth90a,dalessandro91}. It is based on a pump laser beam passing through an optical nonlinear medium (here a cloud of laser-cooled atoms) and being retro-reflected by a plane mirror located at a distance $d$. Any spatial perturbation of the transmitted wavefront due to small fluctuations of the cloud refractive index $n$ along the transverse directions can convert into an intensity perturbation because of diffraction during the free-space propagation to the mirror and back to the cloud. Since the refractive index of the cloud depends on the light intensity $I$, these intensity perturbations further modify the transverse distribution of the refractive index and produce a feedback loop. If this feedback is positive, its effect is to enhance the initial wavefront fluctuations leading to the growth of a macroscopic transverse modulation of both light intensity and refractive index, and to the spontaneous appearance of patterns. The typical length scale $\Lambda$ of the emerging pattern is set by diffraction and can be adjusted by varying the feedback distance~\cite{firth90a,dalessandro91,ciaramella93}: $\Lambda \propto \sqrt{d \lambda}$ (where $\lambda$ is the light wavelength). If the thickness, $L$, of the medium is not negligible compared to the free space propagation distance, diffraction within the medium also plays a role. The qualitative reasoning given above remains valid, though $L$ affects the length scale of the pattern \cite{suppl}.

Our specific nonlinear medium  possesses two distinct mechanisms which can lead to positive feedback. The first mechanism is due to the optomechanical coupling (action of light on atomic external degrees of freedom), while the second one stems from an `electronic' nonlinearity (internal degrees of freedom). Indeed, the index of refraction of our cold atomic cloud can be written in the form
\begin{equation}
n(I,\nabla I) = 1 - \frac{3\lambda^3}{4\pi^2} \, \frac{\delta/\Gamma}{1+ (2\delta/\Gamma)^2} \, \frac{\rho(\nabla I)}{1+s(I)}
\label{index}
\end{equation}
where $\delta = \omega_{\mathrm{l}} -\omega_0$ is the detuning between the laser and atomic frequencies, $\Gamma$ is the natural width, $\nabla I$ the gradient of the intensity and $\rho$ denotes the atomic density. The saturation parameter $s = I/\left[I_{\mathrm{sat}}\left(1+4 \left(\delta/\Gamma\right)^2\right)\right]$, where $I_{\mathrm{sat}}$ is the saturation intensity, quantifies the strength of the atom-light interaction.

The electronic nonlinearity is due to the $1/\left(1+s\right)$ term in Eq.~(\ref{index}), which converts spatial intensity modulations into refractive index modulations via changes of the population of the excited state. It is negligible for $s~\ll~1$ (linear regime) and produces a Kerr-like nonlinear refractive index ($n \simeq n_0 + n_2I$) in the intermediate regime of saturation ($s < 1$). The electronic nonlinearity is self-focusing ($n$ increases with $I$) for $\delta > 0$.

In this work, we focus on the optomechanical nonlinearity which arises from the $\rho(\nabla I)$ term in Eq.~(\ref{index}). The coupling between light intensity and atomic density is due to the dipole force exerted on the cold atoms in the presence of transverse intensity variations, which results in spatial bunching. For $\delta > 0$ the atoms are expelled from high-intensity areas, $\rho$ decreases and $n$ increases when $I$ increases, leading to an effectively self-focusing nonlinearity. We stress that the optomechanical coupling can in principle sustain an instability even in the linear optical regime ($s \ll 1$), provided that the kinetic energy of the atoms is low enough~\cite{tesio13,suppl}.

In our experiment, sketched in Fig.~\ref{fig1}\textbf{B}, $^{87}$Rb atoms are released from a large magneto-optical trap (MOT) at a temperature of 290~$\mu$K. A collimated and linearly-polarized pump beam is retro-reflected by a mirror (M) after passing through the cloud. The full experimental sequence alternates a preparation stage where the atoms are trapped and cooled, and a measurement stage where the trapping lasers and magnetic field of the MOT are switched off, the pump beam is switched on and the images of the transmitted pump beams are acquired.

\begin{figure}
\begin{center}
\includegraphics[width=160mm]{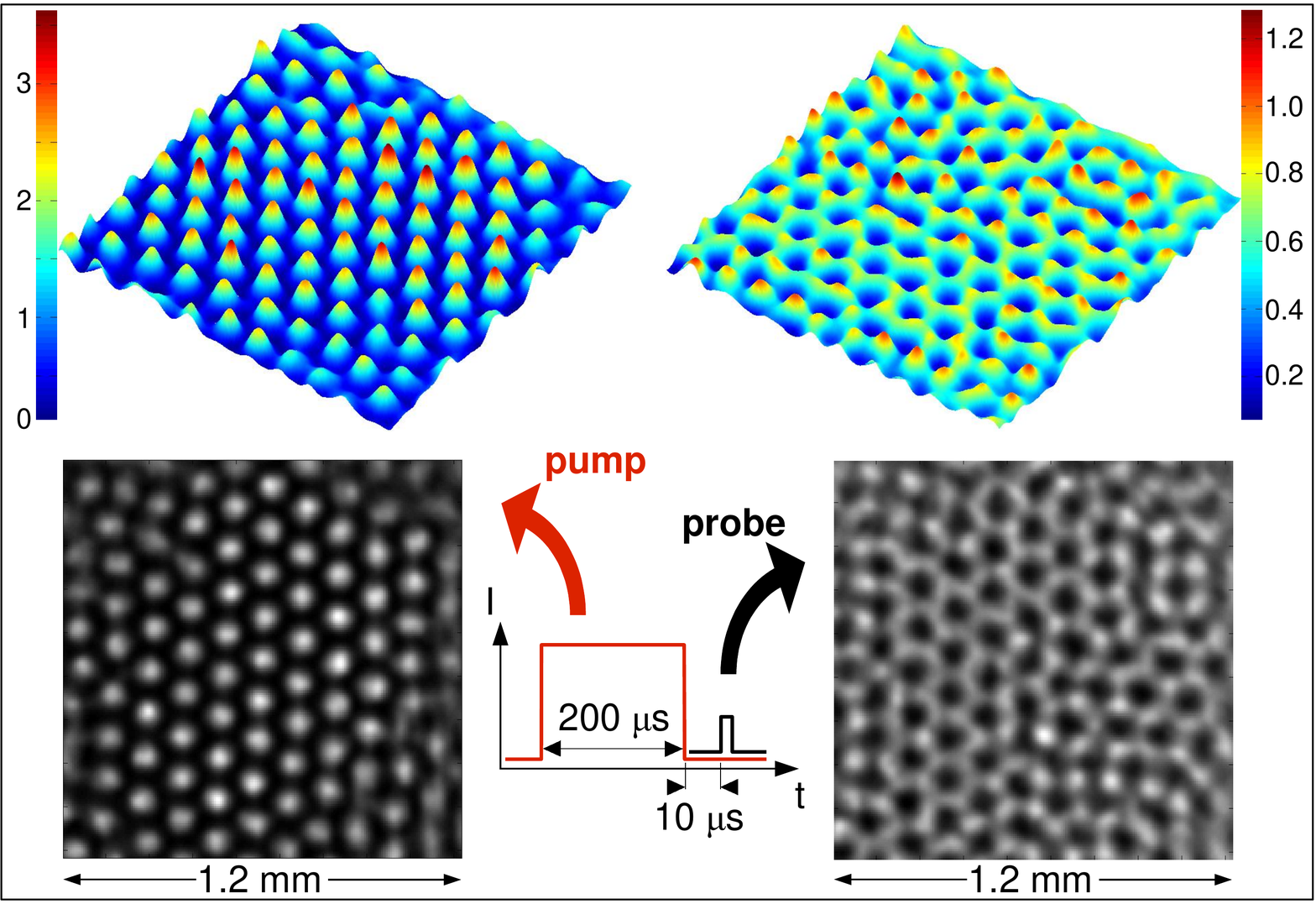}
\caption{Observation of optical patterns and self-structuring of the atomic density. The grey-scale images are single-shot measurements of the transmitted pump (left) and probe (right) beams, with the corresponding 3D representations of the intensity distributions shown above. The pump image is recorded at the end of the $200~\mu$s pulse, with parameters $I = 129\,$mW/cm$^2$, $\delta = +7\Gamma$, and $d=5$~mm. The probe image is recorded $10~\mu$s after the pump has been switched off, and reflects the transverse spatial bunching of the atomic cloud (see text). Both images are normalized to the incident beam intensities (recorded in the absence of cold atoms). The pump transmission exhibits extended, high-contrast hexagonal patterns, while the probe transmission reveals the existence of a complementary honeycomb pattern for the atomic density.}
\label{fig2}
\end{center}
\end{figure}

Typical images of the transmitted pump beam intensity profile exhibit a high-contrast hexagonal pattern (Fig.~\ref{fig2}, left panel), which extends over a large portion of the beam cross-section. The intensity in the bright spots of this pattern is more than three times larger than the peak intensity of the incident pump beam. We typically observe patterns for $3 \Gamma < \delta < 25 \Gamma$, above threshold values for pump intensity and cloud optical density ($OD$)~\cite{suppl}.

\begin{figure}
\begin{center}
\includegraphics[width=120mm]{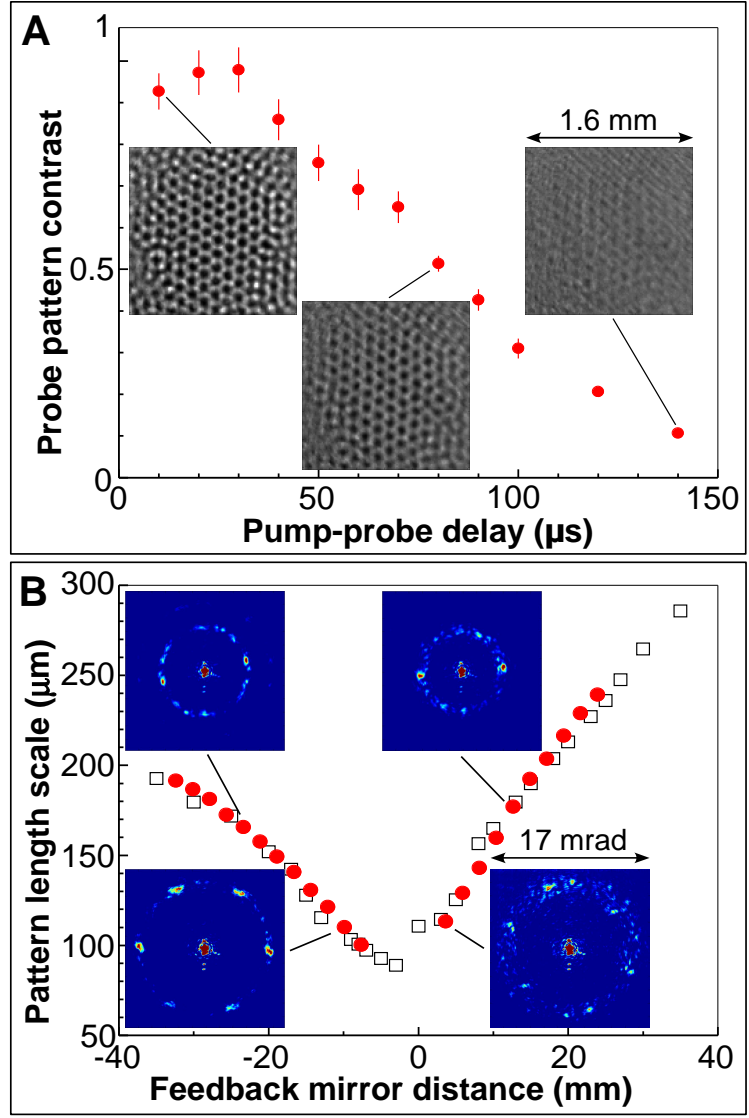}
\caption{(\textbf{A}) Decay of the density pattern (as measured by the probe) as the delay between pump and probe pulses is increased. The parameters are those of Fig.~\ref{fig2}, with a fixed pump duration $\Delta~t~=200~\mu$s. A long decay time scale of $\approx 80~\mu$s (corresponding to a decrease of the contrast by a factor of two) is observed, consistent with the ballistic expansion of the atoms due to their transverse velocity spread. (\textbf{B}) Variation of the pattern length scale $\Lambda$ with the feedback mirror distance $d$. The dots are the experimental data, while the squares are obtained from a model including diffraction and the thickness of the atomic cloud \cite{suppl}. The insets show far-field images used to measure $\Lambda$.}
\label{fig3}
\end{center}
\end{figure}

The tunability of the pattern length scale $\Lambda$ is illustrated in Fig.~\ref{fig3}\textbf{B}. The dependence of the length scale on the mirror distance confirms that the conversion from phase to intensity fluctuations in the feedback loop is the key ingredient for the occurrence of the instability. The experimental results (circles) are in excellent agreement with the prediction from a thick-medium model (squares, see \cite{suppl}).

The pattern in the pump beam is accompanied by a spatial self-structuring of the atomic cloud, as measured by a weak probe beam, polarized orthogonally to the pump and sent through the cloud after the pump pulse (right image of Fig.~\ref{fig2}; details are in `Methods'). This atomic density image displays a high-contrast honeycomb pattern complementary to the hexagonal pattern of the pump, as expected.

A further confirmation of the spatial bunching of the atoms is obtained by measuring the decay of the probe pattern as the time delay between pump and probe pulses is increased (Fig.~\ref{fig3}\textbf{A}). We observe a slow decay (time scale $\approx 80~\mu$s) which excludes the possibility that the pattern observed in the probe profile is due to a transverse modulation of the excited state population (which has an electronic time constant of $\Gamma^{-1}=26$~ns), but is compatible with the free ballistic spreading of the atoms once the pump pulse has ended~\cite{suppl}. Zeeman effects and hyperfine modulations were excluded by further measurements ~\cite{suppl}. This observation constitutes direct evidence of spontaneous transverse organization and symmetry breaking of the atomic density under the action of a single retro-reflected laser beam. The observation of a strong transverse spatial bunching is consistent with the fact that the estimated dipole potential height lies in the hundreds of $\mu$K to mK range, comparable to or larger than the MOT temperature \cite{long_bunch}.
It is also in qualitative agreement with predictions from a model describing the coupled dynamics of the light field and the atomic density~\cite{suppl}.

\begin{figure}
\begin{center}
\includegraphics[width=160mm]{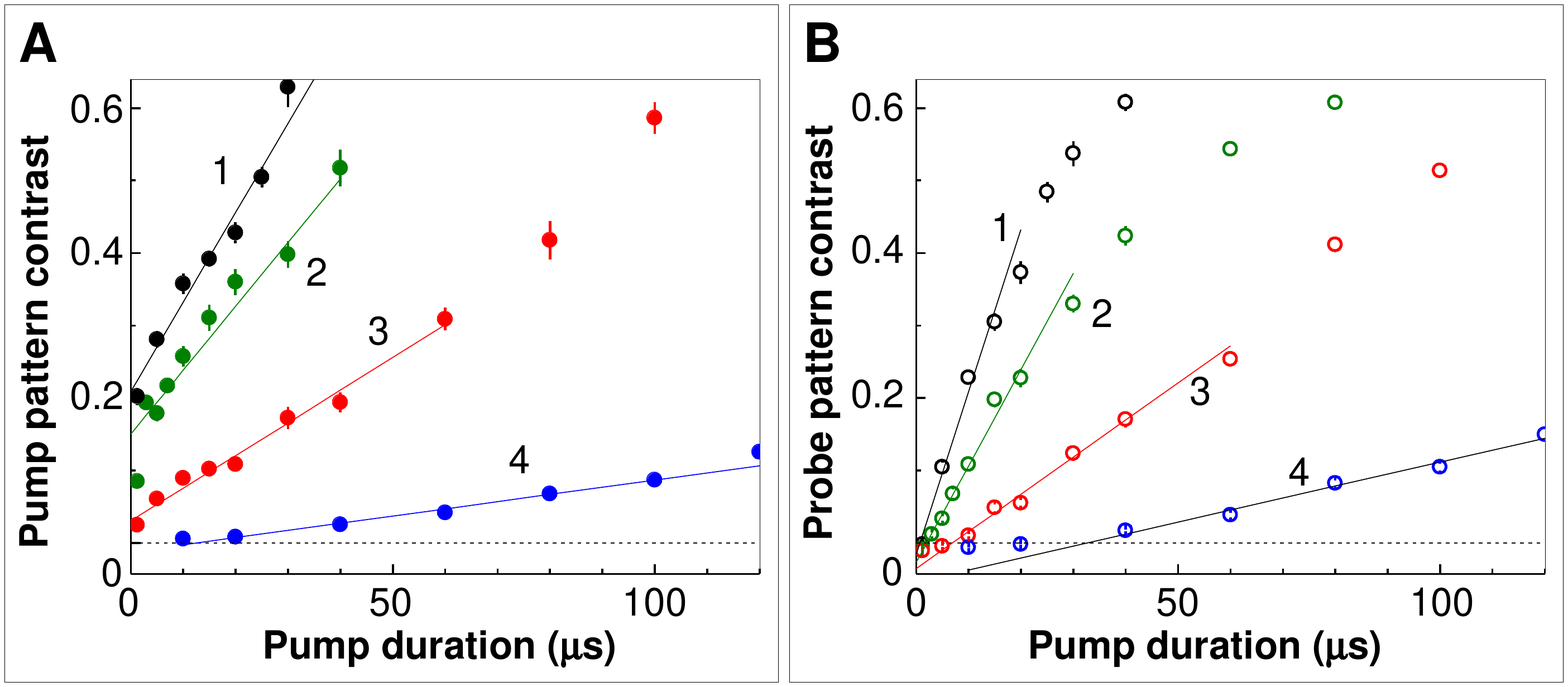}
\caption{Optomechanical vs electronic nonlinearity. We plot the pattern contrast measured in the transmitted pump (\textbf{A}) and probe (\textbf{B}) beams, as a function of the pump duration $\Delta t$ and for different pump intensity levels. The probe pulse is sent $10\,\mu$s after the end of the pump pulse. The pump and probe detunings are respectively $+10\Gamma$ and $-6\Gamma$. The pump intensities are: (1) $I = 636\,$mW/cm$^2$, (2) $I = 404\,$mW/cm$^2$, (3) $I = 217\,$mW/cm$^2$ and (4) $I= 91\,$mW/cm$^2$. The lines are linear fits.}
\label{fig4}
\end{center}
\end{figure}

Simultaneous monitoring of light patterns on the pump and the probe provides insights into the instability mechanisms at work. We measured the evolution of the pattern contrast in the pump and probe as the duration and intensity of the pump pulse was varied (Fig.~\ref{fig4}). The observed increase of contrast with pump duration, on a time scale (several tens of $\mu$s) much larger than the electronic time constant is a clear manifestation of the optomechanical nonlinearity. This interpretation is supported by the comparison of the pump and probe patterns at the shortest pump duration ($1~\mu$s): while the probe pattern is always absent, the pump pattern is on the contrary already sizeable for large enough intensities (see curves (1) and (2) in Fig.~\ref{fig4}). This is interpreted to be a regime in which the electronic nonlinearity alone can trigger pattern formation, i.e.\ curves (1) and (2) are above the electronic instability threshold. Afterwards, the rather `slow' optomechanical nonlinearity leads to atomic bunching, resulting in turn in a significant enhancement of the contrast of the light pattern.

For low pump intensity (curve (4) in Fig.~\ref{fig4}), one observes a co-evolution of both pump and probe patterns, with patterns appearing in the atomic density and light intensity only after a relatively long pump duration. In this regime,  the system is below the electronic instability threshold and the self-organization is clearly dominated by the optomechanical nonlinearity. For the parameters of curve (4), the saturation corresponding to the incident pump is only $s \simeq 0.06$. Pattern formation at even smaller saturations ($s \ll 1$) should be possible \cite{tesio12,tesio13,suppl}, provided that a colder atomic cloud is used, bringing the system even further into the optomechanical instability regime.

In conclusion, we have observed the large-scale transverse self-structuring of a cold atomic medium under the action of a single retro-reflected laser beam, due to optomechanical coupling. Natural extensions of this work facilitating optomechanical feedback will be to decrease the initial temperature of the atomic ensemble or to introduce velocity damping \cite{voncube04,muradyan05,greenberg11,tesio12}. The length scale of the pattern can be easily adjusted via the distance of the feedback mirror to the medium. Spontaneous symmetry breaking in the transverse plane results in hexagons. Close to threshold hexagonal patterns typically possess the intriguing feature that individual constituents of the  pattern can serve as `dissipative solitons' \cite{akhmediev05}.  This indicates that self-sustained localized holes or peaks could be written into the atomic density distribution and erased again `at will' by external control pulse.

The mechanism of self-structuring described here is expected to apply also to instabilities of the electron density in a plasma due to ponderomotive forces \cite{tesio13} and to `soft matter' formed from dielectric beads \cite{smith81,reece07} with optical feedback. Corresponding diffractive instabilities extend to cavity systems with atoms \cite{tesio12} or deformable membranes \cite{ruiz-rivas12} as optomechanical elements.
New intricate features can be expected by the interaction of matter wave nonlinearity, quantum degeneracy and optomechanical structures, if cold atoms are replaced by Bose-Einstein condensates, or in regimes with multiple photon scattering, leading to diffusive transport of light \cite{mendonca12}.

\subsection*{Methods}
The light-atom interaction described in this experiment is based on the $F = 2 \rightarrow F' = 3$ transition of the $D_2$ line of $^{87}$Rb at $780.24$~nm. The natural linewidth of this transition is $\Gamma = 2\pi\times6.067$~MHz and the saturation intensity is $I_{\mathrm{sat}} = 3.58$~mW/cm$^2$ assuming an equal population distribution among Zeeman substates. We stress that the distinction between the
optomechanically triggered and electronically triggered case (see discussion of Fig.~\ref{fig4}) stems from experimental observation of the dynamics, not from an estimate of the saturation level.
All results presented here are obtained with a few percent (compared to the pump light) of repumping light (tuned close to the $F = 1 \rightarrow F' = 2$ transition) copropagating with the pump beam. The atomic cloud has a Gaussian density profile with dimensions (full width at half maximum) of ($10 \times 10 \times 5$)~mm$^3$ (10 mm along the pump propagation axis), and contains $6 \times 10^{10}$ atoms. The peak spatial density is $10^{11}$ cm$^{-3}$ corresponding to an $OD = - \hbox{ln}(T_I) = 150$ (where $T_I$ is the fraction of transmitted light intensity) at line centre (see also \cite{labeyrie11}). The pump beam is directed to the atoms via a monomode polarization-maintaining optical fiber. It is collimated and linearly-polarized (waist $1.6~$mm).  The feedback mirror is located outside the vacuum chamber where the cold atoms are produced. We therefore use a telescope to image the mirror onto a plane located near the cold atomic cloud, producing a `virtual mirror' whose distance $d$ to the sample back end from the sample can be adjusted, from positive to negative values \cite{ciaramella93}. The typical saturation parameter range for the pump is $0.05 < s < 1$. The image plane of the transmitted pump beam is located $\simeq 10$~mm after the cloud centre. For the patterns shown in Figs.~\ref{fig2} and \ref{fig3}A, this corresponds to the feedback beam reinjected into the medium.

It was checked that the instability is not a polarization instability, i.e.\ the spontaneous sidebands have the same polarization as the pump beam. This enables us to introduce a polarizing beam splitter (PBS) in the feedback loop so that we can use an orthogonally polarized probe beam to directly visualize the atomic density self-structuring resulting from the optomechanical instability. The probe beam is sent through the same fibre as the pump, typically $10\,\mu$s after the pump has been switched off. After transmission through the cloud, it is reflected by the PBS located before the mirror, and sent to a CCD (probe image). The probe beam does not reach the mirror and it is short (10~$\mu$s) and weak ($s \ll 1$) and thus does not exert any feedback on the atoms.  Because of the large on-resonance OD of the cloud, standard absorption imaging cannot be employed to measure the transverse density distribution of the atoms. Instead, we take advantage of the dispersive behaviour of the cloud by using a probe detuning of $-~7~\Gamma$ where absorption is reduced. The plane imaged by the probe detection optics is  the same as for the pump beam. The fact that the pump and probe images are complementary illustrates that dispersive imaging is dominating the absorption effect and the resulting pump and probe patterns are compatible with a honeycomb atomic density structure, since light concentrates in regions of high refractive index, which for the negative probe detuning corresponds to regions of high atomic density and for blue detuning to the voids (see Fig.~\ref{fig2}).

\subsection*{Acknowledgements}
The Strathclyde group is grateful for support by the Leverhulme Trust and EPSRC, the collaboration between the two groups is supported by the Royal Society (London). The Sophia Antipolis group acknowledges support from CNRS, UNS, and R\'{e}gion PACA.

\newpage

\subsection*{Supplementary materials}

\section{Details on experiment and analysis}

\textbf{Density pattern vs internal state pattern}

The instability described in this Letter relies on the optomechanical coupling between the light field and the atoms, whose response is well reproduced by a simple 2-level model. However, the complex internal structure of the atoms may also contribute additional nonlinearities. For instance, optical pumping between Zeeman substates is at the heart of the polarization instability observed in Refs.~\cite{gauthier88,petrossian92,grynberg94,aumann97}. Hyperfine pumping has also been shown to play a role in self-lensing~\cite{labeyrie07}. We observe in Fig.~2 patterns in the transmitted probe beam, that we interpret as a direct consequence of spatial self-structuring of the cloud density distribution. To support this claim, we carefully checked alternative origins for the observed pattern, namely transverse inhomogeneous populations of internal states (Zeeman or hyperfine).

First, a J$\to$J+1 transition as investigated here does not possess dark states, which decouple from the light-atom interaction and are responsible for polarization preserving Zeeman patterns \cite{ackemann94,ackemann01} and polarization instabilities \cite{gauthier88,petrossian92,grynberg94,aumann97,ackemann01} described before in hot Na and Rb vapour in the vicinity of the D$_1$-line. Hence Zeeman pumping serving as the basis of a nonlinear effect is not expected.
Second, the formation of high-contrast Zeeman-state gratings is expected to be highly polarization- and magnetic field-dependent since it requires the use of a carefully controlled polarization and the application of a magnetic field of appropriate orientation and strength. On the contrary, in our experiment the patterns are observed to be quite independent of the polarization configuration: we observed patterns for a lin // lin (same linear polarization for input and retro-reflected beams), $\sigma^+ - \sigma^+$ (same circular polarization) and $\sigma^+ - \sigma^-$ (orthogonal circular polarizations) configuration. For the lin // lin configuration, we observed no difference between patterns obtained with a well-compensated magnetic field and with an applied bias field of a few Gauss of various orientations (parallel to the pump propagation axis, parallel to the pump polarization, or orthogonal to both). In addition it was checked that the instability modes have the same polarization as the pump mode, i.e.\ there is no polarization instability. These tests appear to rule out Zeeman induced population patterns.

We also checked the population of the lower $F = 1$ hyperfine state after the pump pulse and found that it is rather small (at most $10\%$) and cannot account for the large modulation observed in the probe transmission.

Finally, the slow decay of the probe pattern with increasing pump-probe delay (see Fig.~3) rules out an occupation of the excited state as the origin of the probe modulation and is compatible with a wash-out of the density pattern due to the velocity distribution of the atoms.

\textbf{Optomechanical vs electronic nonlinearity}

As discussed briefly in the article, the time scales for the growth of the probe pattern (Fig.~4) and its decay (Fig.~3A) indicate that there is an optomechanical instability leading to transverse spatial bunching. As emphasized before and worked out in detail below (Sec.~\ref{sec:model}), this instability can develop even in the absence of any electronic effect, in the linear optical regime. For pump intensities in the few $100~$mW/cm$^2$ range, however, the two instability mechanisms (optomechanical and electronic) coexist and can then be distinguished via their different time scales by varying the pump pulse duration as shown in Fig.~4. Further evidence is provided in Fig.~1S, where we compare the $OD$ thresholds for a short ($1\, \mu$s) and long ($200\,\mu$s) pump pulse. In this measurement, the $OD$ of the cloud is varied by tuning the time delay between the extinction of the MOT and the pump pulse. Because of the cloud's ballistic expansion after release from the MOT, the $OD$ monotonically decreases with increasing delay. The pump intensity is $487\,$mW/cm$^2$, and its detuning $+6\Gamma$. For the short pulse (electronic instability), the
contrast of the pump pattern increases above a threshold $OD\approx 89$. In the case of the longer pulse, due to the optomechanical mechanism, the threshold $OD$ is considerably reduced to a value around $19$.

\textbf{Possible influence of longitudinal bunching and thick medium} \label{sec:long}

For parallel incident and reflected pump polarizations, a standing wave establishes inside the cloud possibly leading to longitudinal spatial bunching of the atoms (see, e.g., \cite{greenberg11} for counter-propagating pump beams at $\delta <0$). Qualitatively, however, longitudinal bunching seems to play a rather unimportant role in our experiment since similar patterns are observed for parallel polarizations (presence of a standing wave) and orthogonal polarizations (no standing wave in intensity, but in polarization state), either linear or circular. A possible explanation is that the retro-reflected beam acts as a blue-detuned longitudinal molasses with negative friction, which increases the kinetic energy of the atoms and prevents their confinement in the longitudinal standing wave. This negative friction has no impact in the transverse directions, where only regular heating due to pump photon scattering occurs. The time scales for the growth of the probe pattern (Fig.~4) and its decay (Fig.~3A) indicate that transport processes take place on the transverse length scale of 100~$\mu$m and not on wavelength scales (0.4~$\mu$m), where one expects times in the sub-microsecond to 10~$\mu$s range (e.g.\ an atom at the 1D thermal speed of 0.17~m/s would traverse 0.4~$\mu$m in 2.3~$\mu$s). This hypothesis will be investigated in future work. We will develop in Sec.~\ref{sec:model} the simplest theory which does not take into account a wavelength-scale density grating and will find a good agreement of thresholds with the experiment further supporting that the principal mechanism is captured by the model. We mention that the same assumption was made in the proposal for counterpropagating beams in \cite{muradyan05}.

Fig.~3B clearly establishes a dependence of the pattern length scale $\Lambda$ on mirror distance $d$, but a careful analysis shows that it is only qualitatively described by a theory in which the medium is diffractively thin and all diffraction takes place in the vacuum feedback loop, since in the present experiment the mirror distance is comparable to the medium thickness ($\simeq~1~$cm). The effects of medium thickness are studied here by including mirror feedback in a model of counter-propagation instabilities in a Kerr slab \cite{Geddes94}, i.e.\ a thick medium with $n=n_0+n_2I$ where the $n_2$ might originate from any nonlinearity (here optomechanical or electronic). This changes the boundary conditions on the forward and backward beams and hence modifies the growth condition for a non-trivial solution of the linearized equations for transverse perturbations of wavenumber $q$. This condition determines the threshold for instability of the homogeneous solution, and is typically an undulatory function of $q$. Its minima correspond to `modes' $q_{\rm min}$ with locally minimum threshold and in our model depend on the mirror distance $d$ as well as on the thickness and other parameters of the medium.

Since $d$ enters only through the differential phase shift (proportional to $q^2 d$) between the on-axis pump beam and the $q$-sideband, the theory allows negative values of $d$. This can be achieved in the experiment by imaging the feedback mirror to a plane inside, or even beyond, the atom cloud. As shown in Fig.~3B, the dependence on $d$ for the lowest-wavenumber mode (squares) is in good agreement with our experimental data (circles) across a broad range of positive and negative mirror distances. In this figure the zero of the $d$-axis corresponds to the centre of the nonlinear medium. This good agreement confirms that the cause of the instability is the conversion of phase to intensity modulation in the feedback loop. The tunability of $\Lambda$ is a specific feature of the single mirror-feedback scheme, absent from the counter-propagating scheme with two input beams and no optical feedback~\cite{muradyan05, Geddes94, greenberg11}.

\begin{figure}
\begin{center}
\renewcommand{\figurename}{Fig S}
\includegraphics[width=120mm]{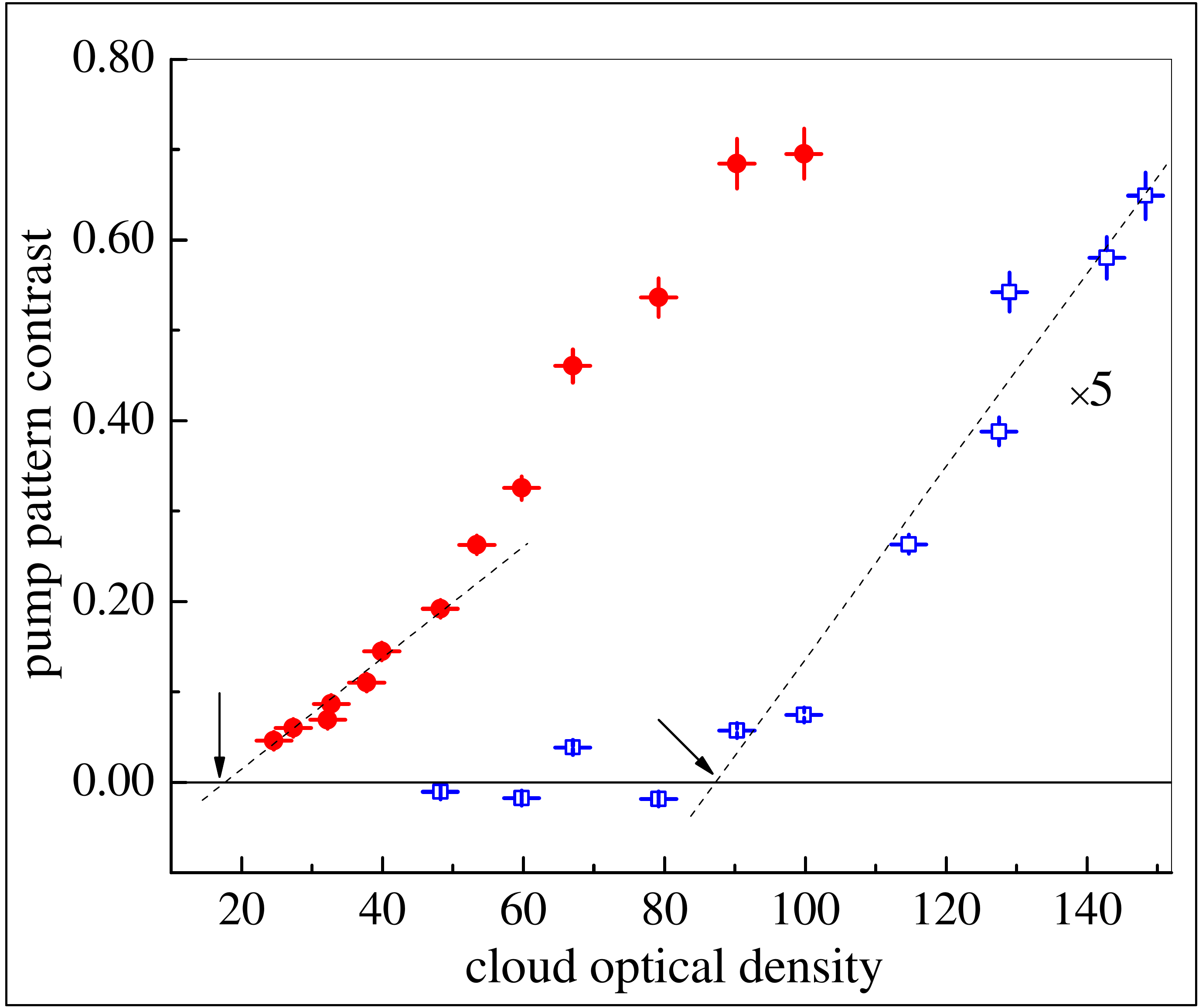}
\caption{Instability threshold behavior versus cloud optical density. We compare the measured evolution of the pattern
contrast with $OD$ for two pump pulses: a short pulse ($1\,\mu$s, blue squares), and a long one ($200\,\mu$s, red circles). In the latter case, the optomechanical mechanism strongly reduces the threshold $OD$. The pump parameters are: $\delta=+6\Gamma$, $I = 487\,$mW/cm$^2$.}
\label{fig1S}
\end{center}
\end{figure}

\section{Theoretical model and numerical results}\label{sec:model}

We have undertaken a theoretical investigation of the optomechanically-driven instability (details can be found in \cite{tesio13}). Previous studies of optomechanical instabilities assumed strong velocity damping due to the presence of optical molasses~\cite{muradyan05,tesio12}. However, our experiment showed spontaneous symmetry breaking in the absence of such damping, with the MOT beams switched off during the interaction. We then describe the atomic dynamics in terms of a damping-free, collision-less Boltzmann equation with a nonlinear term driven by the dipole force. For simplicity, we focus our attention on the optomechanical coupling only and neglect the electronic nonlinearity. Then the state of the cloud is completely described by its phase-space distribution function $f=f(\mathbf x,\mathbf v,t)$ of the cloud (with $\mathbf x$ and $\mathbf v$ position and velocity vectors in the plane transverse to the field propagation). Its equation of motion is
\begin{equation}
\frac{\partial f}{\partial t} + \mathbf v\cdot\frac{\partial f}{\partial \mathbf x} + \frac{\mathbf F_{\textrm{dip}}}{M}\cdot\frac{\partial f}{\partial \mathbf v}=0\label{eq:vlasov} \, .
%\label{eq:system}
\end{equation}
Here  $M$ is the atomic mass and $\mathbf F_\textrm{dip} = -\partial_\mathbf xU_\textrm{dip}$ the dipole force with $U_\textrm{dip}=(\hbar\delta/2)\log(1+s(\mathbf x,t))$, and $s$ is the saturation parameter introduced in the article. The spatial density $\rho(\mathbf x,t)$ is obtained by integrating $f$ over the entire velocity space, with the normalization chosen so that the spatially homogeneous solution corresponds to $\rho=1$. The saturation parameter $s$ is given by the sum of the intensities of the forward field $g_F$ and the backward field $g_B$  (neglecting longitudinal interference effects), suitably normalized to give $s=|g_F|^2+|g_B|^2$. As $g_B$ depends on the density $\rho$ (see below), this closes the feedback loop.

We neglect diffraction effects inside the cloud (thin medium approximation), and absorption by the cloud.
Under these assumptions, the interaction between a forward pump field of amplitude $g_F$ and the cloud of laser-cooled two-level atoms is described by the following equation
\begin{equation}
\frac{\partial g_F}{\partial z} = i\frac{OD\,\delta/\Gamma}{L\,\left(1+4(\delta/\Gamma)^2\right)}\rho \,g_F\label{eq:gF} \, ,
\end{equation}
where $L$ is the medium thickness. To obtain $g_B$, we first integrate Eq.~(\ref{eq:gF}) under the assumption of a longitudinally homogeneous $\rho$ and obtain the transmitted field $g_T$ at the exit facet of the medium. The free-space propagation to the mirror (distance $d$) and back can be solved exactly in Fourier space, and the backward field before re-entrance in the medium is given by:
\begin{equation}
g_B(q)= \sqrt R\,e^{i d q^2 / k_0} g_T(q)\, , \label{eq:gB}
\end{equation}
where $R$ denotes the mirror reflectivity.

The spatially homogeneous solution of the system is given by $f=f_0(\mathbf v)$, $s=(1+R)|g_F|^2$, where $f_0(\mathbf v)$ denotes the initial velocity distribution of the gas. We introduce perturbations in the distribution function and the backward field as $f=f_0+f_1(\mathbf x,t)$, $g_B=g_B^{(0)}[1+b_1(\mathbf x,t)]$, and linearize Eqs.~(\ref{eq:vlasov}), (\ref{eq:gF}), (\ref{eq:gB}) about the homogeneous solution. The instability threshold is then determined using linear stability analysis. For a Maxwell-Boltzmann $f_0(\mathbf v)$ characterized by a temperature $T$ a threshold condition for the injected intensity is found in terms of the relevant parameters ($\delta$, $OD$, $T$) as:
\begin{equation}
|g_F^{\textrm{th}}(q)|^2=\left[\frac{\hbar\delta}{2k_BT}\frac{OD}{1+(2\delta/\Gamma)^2}\frac{2\delta}{\Gamma}R\sin\left(dq^2/k_0\right)-(1+R)\right]^{-1}\, ,
\label{eq:threshold}
\end{equation}
where $\hbar$ is the reduced Planck's constant and $k_B$ is Boltzmann's constant.

Eq.~(\ref{eq:threshold}) shows that an optomechanical instability can occur even in the absence of electronic nonlinearity, and identifies the intensity threshold as a function of all the parameters. However, this requires the expression in the bracket to be positive, which restricts the parameter space where patterns can be observed.

As expected, lower temperatures result in lower intensity and $OD$ thresholds for the instability. We also remark that Eq.~\ref{eq:threshold} can be satisfied only for wavenumbers such that $\sin\left(dq^2/k_0\right)>0$, with the most unstable wavenumber $q_\textrm c=\sqrt{\pi k_0/2d}$ identified by the condition $\sin\left(dq_\textrm c^2/k_0\right)=1$. The spatial scale of the emerging pattern is given by $\Lambda=2\pi/q_\textrm c$ and agrees quite well with the experimental observations (see Sec.~\ref{sec:long} for the quantitative correction due to the thickness of the medium).

The variation of the optomechanical threshold with pattern wavenumber is shown in Fig.~2S\textbf{A} for parameters close to our experimental situation. It is remarkable to see that in spite of the strong approximations, the numerical model predicts the lowest threshold at around $|g_F|^2=0.03$ in Fig.~2S, i.e.\ well within the same order of magnitude of that observed in the experiment (curve (4) in Fig.~4 corresponds to $|g_F|^2\approx 0.06$). For the parameters of Fig.~1S, the predicted threshold in optical density is slightly larger than 12, i.e.\ also in good agreement with the experimental observation of about 19. Some increase in threshold can be expected since the feedback strength is reduced by the residual absorption and losses in the optical components in the feedback loop.

To study pattern formation above the instability threshold, we performed numerical simulations of the system of Eqs.~(\ref{eq:vlasov}), (\ref{eq:gF}), (\ref{eq:gB})  in one transverse dimension, as this is much less demanding computationally and has little impact on the physics of the instability in our situation. Fig.~2S\textbf{B} shows an example of the distributions of the re-entrant light intensity (solid red curve) and atomic density (dashed blue curve) obtained with periodic boundary conditions over a spatial domain of $7\Lambda$. Complementary light and density patterns are obtained, very similar to what is observed in the experiment. We stress that, since electronic effects are completely absent in this model, these structures spontaneously form solely as a result of the optomechanical coupling. In two dimensions, hexagons were previously found in a unidirectional cavity model as the result of optomechanical symmetry breaking instabilities for the case where strong damping is provided by optical molasses~\cite{tesio12}.

\begin{figure}
\begin{center}
\renewcommand{\figurename}{Fig S}
\includegraphics[width=160mm]{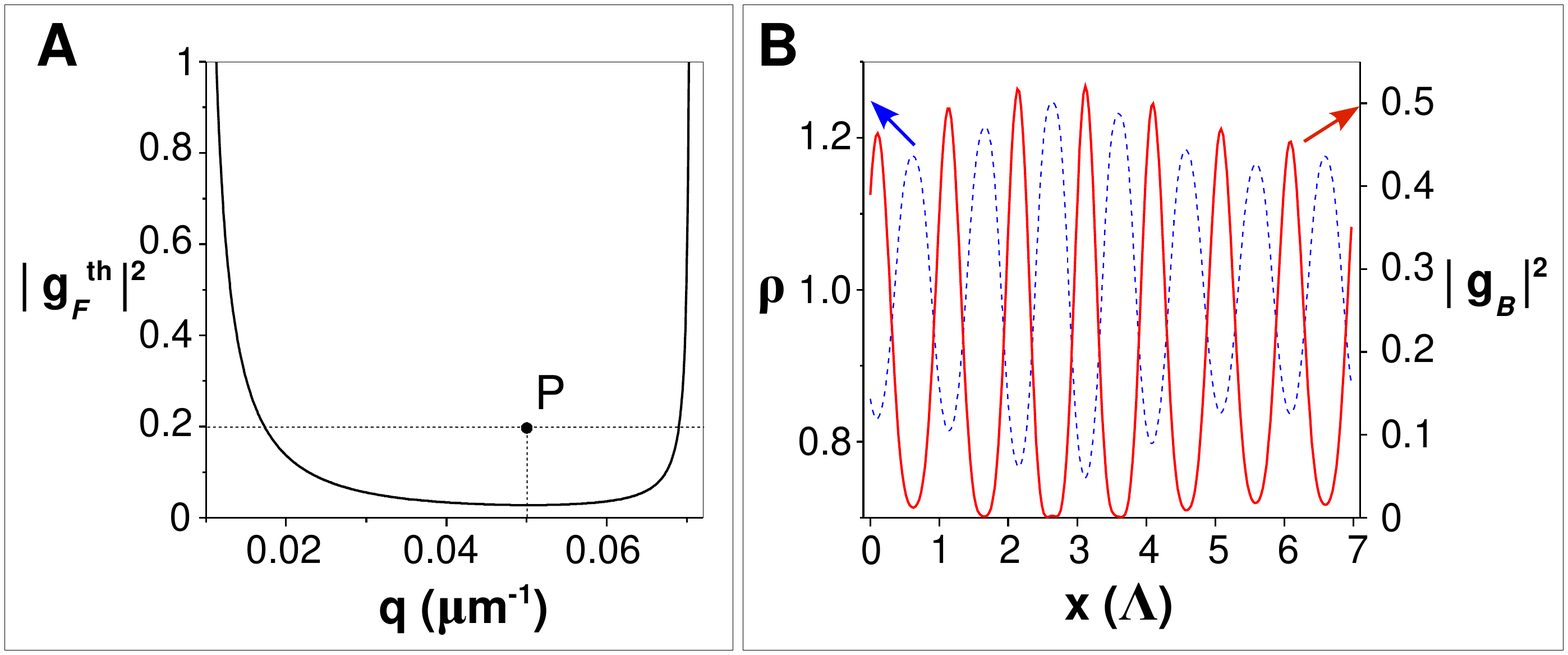}
\caption{Optomechanical instability: theoretical results. (\textbf{A}) shows the intensity threshold vs pattern wavenumber curve obtained from Eq.~(\ref{eq:threshold}). The parameters are: $R=1$, $\delta=+10\,\Gamma$, $OD=150$, $T=290\,\mu K$, $d=5\,$mm. (\textbf{B}) Numerical simulations above the instability threshold (as indicated by point P in (\textbf{A})) showing the transverse intensity distribution of the retro-reflected beam (solid red curve) and the modulation of the atomic density (dashed blue curve). The simulation is implemented over a spatial domain of $7\Lambda$.}
\label{fig2S}
\end{center}
\end{figure}


\begin{thebibliography}{10}
\bibitem{notegreenberg11}
\bibinfo{note}{Indications of transverse pattern formation in a very elongated
  cloud with two counter-propagating pumps along it main axis were reported in
  \cite{greenberg11}. The small transverse extension of the medium however
  precludes the appearance of large extended patterns.}
\end{thebibliography}
\end{document}